\documentclass[twocolumn,showpacs,amsmath,amssymb,superscriptaddress]{revtex4-1}

\usepackage{epsfig,psfrag}
\usepackage{dcolumn}
\usepackage{bm}
\usepackage{graphicx}
\usepackage{color}
\newcommand{\beq}{\begin{equation}}
\newcommand{\eeq}{\end{equation}}
\newcommand{\beqr}{\begin{eqnarray}}
\newcommand{\eeqr}{\end{eqnarray}}

\newcommand{\rb}{{\bar{\rho}}}

\def\tv{{\tilde{v}}}

\def\hi{{\hat i}}
\def\hj{{\hat j}}

\def\bB{{\mathbf B}}

\def\bq{{\mathbf q}}

\def\br{{\mathbf r}}
\def\etab{{\mathbf\eta}}

\def\bQ{{\mathbf Q}}
\def\bK{{\mathbf K}}

\def\bB{{\mathbf B}}

\def\bPi{{\mathbf \Pi}}

\def\bR{{\mathbf R}}

\def\half{{1\over2}}
\def\third{{1\over3}}

\def\eqa{\begin{eqnarray}}
\def\eea{\end{eqnarray}}

\def\pr{Phys. Rev.}
\def\prx{Phys. Rev. X}

\begin{document}

\title{Hamiltonian Theory of Anisotropic Fractional Quantum Hall States}
\author{Ganpathy Murthy}
\affiliation{Department of Physics and Astronomy,
University of Kentucky, Lexington KY 40506-0055}
\date{\today}
\begin{abstract}
Rotationally invariant fractional quantum Hall (FQH) states have long been
understood in terms of composite bosons or composite fermions. Recent
investigations of both incompressible and compressible states in
highly tilted fields, which renders them anisotropic, have revealed
puzzling features which have so far defied quantitative explanation.
The author's work with R. Shankar in constructing and analyzing an
operator-based theory in the rotationally invariant FQHE is
generalized here to the anisotropic case. We compute the effective
anisotropies of many principal fraction states in the lowest and the
first Landau levels and find good agreement with previous theoretical
results. We compare the effective anisotropy in a model potential with
finite sample thickness and find good agreement with experimental
results.
\end{abstract}
\vskip 1cm 
\maketitle

Our understanding of rotationally invariant FQH
states\cite{QH-reviews,laugh} is based on highly accurate variational
wavefunctions\cite{laugh,jain}, which rely on the idea of
particle-flux composites. For fractions of the form
$\nu=\frac{1}{2s+1}$ ($s$ integer) the Laughlin
wavefunctions\cite{laugh} can be understood as ground states of
composite bosons\cite{zhk} (electrons carrying $2p+1$ units of
statistical flux). A unified way to understand all principal fractions
of the form $\frac{p}{2ps+1}$ is  the composite fermion (CF) picture
of Jain\cite{jain}, in which the CF is a composite of an electron and $2s$ quanta
of statistical flux. The CFs see an effective field just right to fill
$p$ CF-Landau levels (CFLLs), thus naturally forming an incompressible
state. CFs can be realized in a field theoretic form by
Chern-Simons theory\cite{lopez-fradkin}, which has had many successes,
notably in the half-filled Landau level\cite{HLR} where the CFs form a
Fermi sea.

Most experimental systems are not rotationally invariant. When the
band mass tensor is isotropic and the magnetic length is much larger
than the lattice spacing, rotational invariance is a good
approximation. Tilting the sample while keeping the filling fixed has
many effects, among which is an induced anisotropy of the band mass
tensor\cite{Kamburov-zero-B}.  Recently, anisotropic transport has
been observed in strongly correlated states in the quantum Hall
regime\cite{gokmen, xia-73,Kamburov1} under strong tilted fields. It
is found that the CF anisotropy is considerably smaller than the
electronic one\cite{Kamburov1}, and at $7/3$, a peculiar
low-temperature\cite{xia-73} state with quantized Hall resistance, but
highly anisotropic longitudinal resistance ($R_{yy}$ vanishing, but
$R_{xx}$ seemingly finite) is seen. Taking steps towards understanding
such states quantitatively will be the focus of this paper. Our
approach will also apply to anisotropic cousins of the recently
discovered FQH-like states in Chern
bands\cite{chern} and two-dimensional time-reversal-invariant
topological insulators (2DTIs)\cite{2dti,2dti-numerics}.

Haldane\cite{Haldane-geo} noted recently that there is an intrinsic
geometry to the FQH regime. Start with an anisotropic band Hamiltonian
in a uniform perpendicular $\bB$ field
\beq H_0=\frac{1}{2m}
\bigg(\frac{\Pi_{ex}^2}{\alpha^2}+\Pi_{ey}^2\alpha^2\bigg) 
\label{h0}\eeq

where the subscript $e$ reminds us that the coordinate
$\br_e=x_e\hi+y_e\hj$ and mechanical momentum $\bPi_e$ are electronic
operators, and $\alpha$ is our electronic anisotropy parameter. The
cyclotron frequency is $\omega_c=\frac{eB}{m}$ and the magnetic length
is $l=\sqrt{\frac{h}{eB}}$. The electronic Hilbert space can be
decomposed into two ``cyclotron'' coordinates ($a,b=x,y$ and
$\epsilon_{ab}$ is the two-dimensional antisymmetric symbol)
$\eta_{ea}=-l^2\epsilon_{ab} \Pi_b$ and two guiding center coordinates
$R_{ea}=r_{ea}-\eta_{ea}$. These two sets have the commutation
relations $[\eta_{ex},\eta_{ey}]=il^2=-[R_{ex},R_{ey}]$, and
$[\eta_{ea},R_{eb}]=0$.

Letting $i$ index particle number, the density operator in first quantization is 

\beq
\rho_e(\bq)=\sum\limits_{i}e^{i\bq\cdot\etab_{ei}}e^{i\bq\cdot\bR_{ei}}
\eeq

When projected to the $n^{th}$ LL, this density becomes
\beq
e^{-|z|^2/2}L_n(|z|^2)\sum\limits_{i}e^{i\bq\cdot\bR_{ei}}=e^{-|z|^2/2}L_n(|z|^2)\rb_e(\bq)
\label{proj-of-rhoe}\eeq

where $z=i\frac{l}{\sqrt{2}}\big(\frac{q_x}{\alpha}+iq_y\alpha\big)$, and
$L_n$ is the $n^{th}$ Laguerre polynomial. Here the projected guiding center density
$\rb_e(\bq)$ is also the operator generating magnetic translations in a
given LL, and obeys the GMP algebra (named in honor of Girvin,
MacDonald, and Platzman\cite{GMP})

\beq
[\rb_e(\bq),\rb_e(\bq')]=2i\sin\bigg(i\epsilon_{ab}q_aq'_b\frac{l^2}{2}\bigg)\rb_e(\bq+\bq')
\label{GMP}\eeq

Since the kinetic energy is degenerate, the Hamiltonian when projected to the $n^{th}$ Landau level is 
\beq
\bar{H}=\half\int\frac{d^2q}{(2\pi)^2} \tv(q):\rb_e(\bq)\rb_e(-\bq):
\label{hproj}\eeq
where the effective electron interaction is $\tv(\bq)=v(q)e^{-|z|^2/2}L_n(|z|^2)$.

Haldane pointed out\cite{Haldane-geo} that the effective anisotropy of
the FQH state would be a compromise between the anisotropies of the
band mass and the interaction anisotropy (here assumed to be 1). 

Very early work on anisotropic states\cite{early-aniso} focussed on
spontaneous breaking of rotational symmetry. An application of
Chern-Simons theory predicted that the electronic and CF anisotropies
should be identical\cite{balagurov} at $\nu=\half$, in contradiction with
experiment\cite{Kamburov1}. More recent theoretical work on
anisotropic FQH states has been mainly numerical, involving comparing
ground states obtained by exact diagonalization with model
wavefunctions\cite{Model-wfs,variational-Haldane,numerics}. The
special case of a gaussian interaction in the LLL can be analyzed
exactly\cite{kunyang}, but unfortunately this cannot be extended to
realistic interactions, or other Landau levels.

In this paper we analytically compute ${\alpha^{CF}}$ as a function of $\alpha$
given the Landau level one is projecting to, the fraction, and the
form of the interelectron interaction. We will test this approach by
comparing to the numerical results\cite{variational-Haldane}, and find
good agreement. We will compare the effects of band mass
anisotropy in the zeroth and the first Landau levels for the Coulomb
interaction, and consider the effects of finite sample
thickness. Finally, we will compare to experimental results on the
relation between ${\alpha^{CF}}$ and $\alpha$ for the half-filled Landau level\cite{Kamburov1},
and see agreement for reasonable sample thickness.

The method we use is a generalization of the approach developed by
R. Shankar and the author more than a decade
ago\cite{prl-us,rmp-us}. This approach starts with a rewriting of the
electronic Hamiltonian in a given Landau level in an enlarged Hilbert
space spanned by operators obeying CF commutation relations. The
spurious degrees of freedom have to ultimately be projected out in a
suitable manner\cite{conserving,resp-fns-us}. While approximate, this
approach allows us to compute various quantities which are difficult,
if not impossible, in conventional wavefunction or exact
diagonalization treatments, such as response
functions\cite{resp-fns-us}, the effects of nonzero
temperature\cite{finiteTus}, or 
disorder\cite{disorder-us}.

In a particular LL, the electronic degrees of freedom are the
guiding centers $R_{ex},\ R_{ey}$. To make a complete fermionic
Hilbert space we add by hand two pseudovortex degrees of freedom per
electron $R_{vx},\ R_{vy}$, so called because they have the commutation relations of a double vortex
$[R_{vx},R_{vy}]=\frac{il^2}{2\nu}=\frac{il^2}{c^2}$
which defines $c=\sqrt{2\nu}$ for the case of two flux quanta
``attached''. Since they are unphysical degrees of freedom, they
commute with $\bR_e$: $[R_{ea},R_{vb}]=0$. Now we re-express $\bR_e,\ \bR_v$ in terms of
CF coordinates and velocities. Note that the CF degrees of
freedom have no subscripts, and obey the commutation relations
$[r_{xi},r_{yi}]=0$, $[r_{ai},\Pi_{bi}]= i\delta_{ab}$, and most
importantly,
$[\Pi_{xi},\Pi_{yi}]=\frac{i}{(l^*)^2}=\frac{i(1-c^2)}{l^2}$. The last
relation shows that the CF's move in a reduced field.
Define the
CF-cyclotron variables by $\eta_a=-(l^*)^2\epsilon_{ab}\Pi_b$, and the
CF guiding center variables by $R_a=r_a-\eta_a$. In the isotropic FQH regime, we made the identification
\beq
\bR_e=\bR+c\etab\ \ \ \ \ \ \ \bR_v=\bR+\etab/c
\eeq
We expressed the projected density $\rb(\bq)$ in terms of CF
variables and proceeded to find a Hartree-Fock ground state (CFHF for short). For
$\nu=\frac{p}{2p+1}$ the HF ground state was just $p$ filled CF-Landau
levels (CFLLs).

The CF-substitution for the anisotropic problem is similar.  We leave
the expression for $\bR_v$ unchanged but make the expression for
$\bR_e$ anisotropic:
\beq
R_{ex}=(R_x+c\eta_x)/{\alpha^{CF}}:\ \ \ \ R_{ey}=(R_y+c\eta_y){\alpha^{CF}}
\eeq

Note that the $\bR_e$ and thus the Hamiltonian still commute with
$\bR_v$, and the constraint structure is unaltered. Now we can express
the Hamiltonian in terms of CF variables and proceed as in the
isotropic case.

With an eye to generalizing this approach to anisotropic states in
Chern bands and 2DTIs\cite{TI-us} we will find it convenient to use crystal
momenta in a Brillouin Zone in the given Landau level. For a principal
fraction $\frac{p}{2p+1}$ it is now straightforward to rescale the
momenta ($Q_x=q_x/\alpha^{CF},\ Q_y=q_y\alpha^{CF}$), choose a unit
cell penetrated by an integer number of effective flux quanta
($a(2p+1)$ in the $x$-direction and $a$ in the $y$-direction, where
$a^2=2\pi l^2$), and thus define a single-CF Brillouin Zone (details
can be found in ref. \cite{TI-us}) with canonical fermion operators
$d^{\dagger}_{n\bK},\ d_{n\bK}$, in terms of which the electron
guiding center density and Hamiltonian are
\begin{figure}
\includegraphics[scale=0.35]{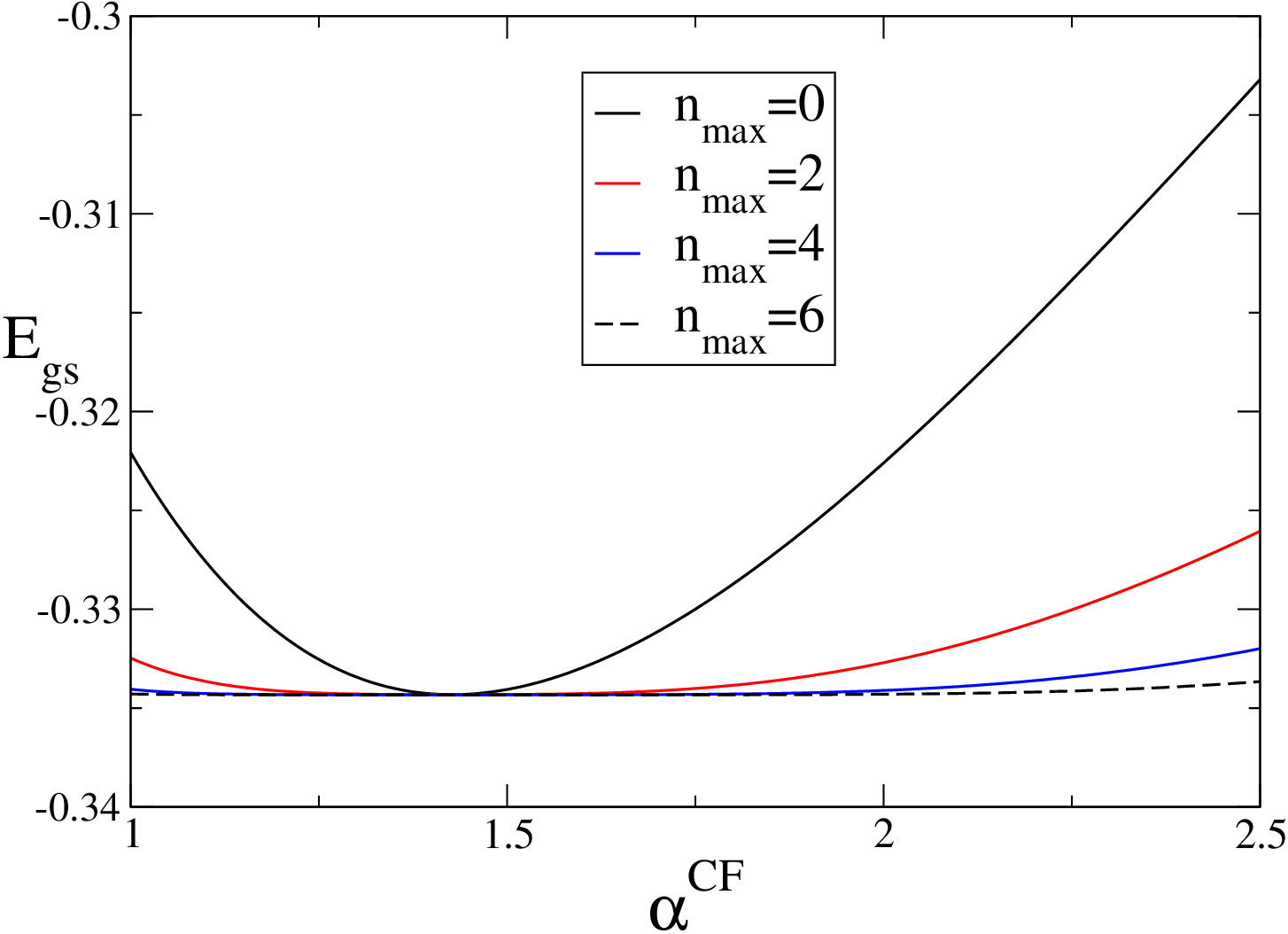}
\caption{The CFHF energy per particle at $\third$ in the LLL for $\alpha=2$  
versus ${\alpha^{CF}}$ for different numbers of CFLLs kept
($n_{max}$). The interaction is pure Coulomb. }
\label{egs13vsllmax}
\end{figure}
\beqr
\rb_e(\bQ)=&\sum\limits_{\bK\in BZ, n',n}\rho_{n'n}(\bQ)e^{i\Phi(\bQ,\bK)}d^{\dagger}_{n'[\bK+\bQ]}d_{n\bK}\\
H=&\frac{1}{2L^2}\sum\limits_{\bQ}\tv(\bQ) :\rb_e(\bQ)\rb_e(-\bQ):
\eeqr
Here $[\bK+\bQ]$ belongs to the BZ, and is defined by $\bK+\bQ=[\bK+\bQ]+\frac{2\pi}{a} N_y(\bQ,\bK)+\frac{2\pi}{a(2p+1)} N_x(\bQ,\bK)$, and 
$\Phi(\bQ,\bK)=\frac{2p+1}{a^2}\big(-(K_x+Q_x)N_y(\bQ,\bK)+\frac{Q_x}{2\pi}(K_y+\frac{Q_y}{2})\big)$. 

To define the matrix elements $\rho_{mn}(\bQ)$ compactly we need $Z=\frac{cl^*}{\sqrt{2}}(Q_x+iQ_y)=|Z|e^{i\theta}$,
in terms of which
\beqr
\rho_{n'n}(\bQ)=&(-1)^{min(n',n)+n}\sqrt{\frac{min(n',n)!}{max(n',n)!}}e^{-|Z|^2/2}\nonumber \\
&L_{min(n',n)}^{|n'-n|}(|Z|^2)|Z|^{|n'-n|} e^{i(n'-n)(\theta-\pi/2)}
\eeqr

Translationally invariant CFHF states for $\nu=\frac{p}{2p+1}$ 
correspond to averages
\beq
\langle d^{\dagger}_{m\bK'}d_{n\bK}\rangle=\delta_{\bK,\bK'}\Delta_{mn}
\eeq
where $\Delta_{mn}$ are independent of $\bK$. 
The $\nu=\half$ state will be treated as the $p\to\infty$ limit.

Two important points need to be noted: (i) The CFHF energy is
variational. The Hamiltonian commutes with all the $\bR_{vi}$, so the
exact ground state in the enlarged Hilbert space must be a direct
product of the exact ground state in the $\bR_e$ sector and an
arbitrary state in the $\bR_v$ sector. Since the Hamiltonian is
independent of $\bR_v$ the exact ground state energy in the enlarged
space is the same as that in the electronic space. (ii) For
incompressible fractions, every set of $\Delta_{mn}$ defines a Slater
determinant state, so for fixed $\alpha,{\alpha^{CF}},\nu$ there is
still a lot of freedom. For a given $\alpha,\ {\alpha^{CF}}$, we
iterate the HF procedure till we have a self-consistent set of
$\Delta_{mn}$.

These points are illustrated in Fig. \ref{egs13vsllmax}, which shows
the ground state energy per particle for $\third$ at an electronic
anisotropy of $\alpha=2$ as a function of the assumed CF anisotropy
${\alpha^{CF}}$. The different curves represent different numbers of
CFLL's kept in the variational calculation. {\it All the results we
present are for the Coulomb interaction $v(q)=2\pi e^2/\varepsilon q$
unless stated otherwise.}
\begin{figure}
\includegraphics[scale=0.35]{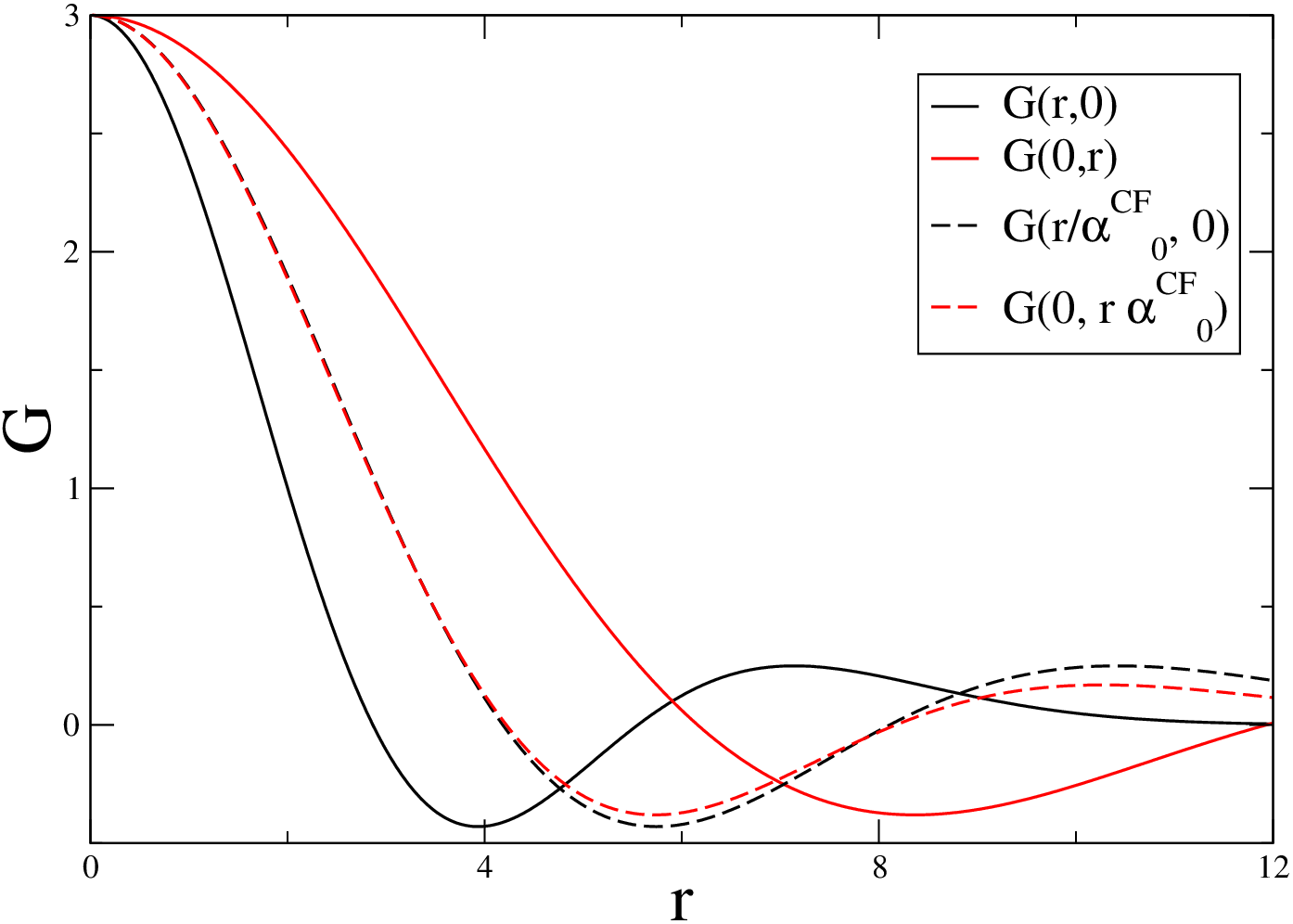}
\caption{The one-body correlator in the $x$ and $y$-directions at $\nu=\frac{3}{7}$ in 
the LLL for $\alpha=2$.  The solid lines are the unscaled correlators,
while the dashed lines are the corresponding scaled correlators for
$\alpha^{CF}_0=1.46$. }
\label{correlator37}
\end{figure}
As can be seen, as $n_{max}$ increases the energy becomes extremely
flat, i.e., nearly independent of ${\alpha^{CF}}$. This is because the
full set of CFLLs at $\alpha^{CF}_1$ is related to the full set at
$\alpha^{CF}_2$ by a unitary transformation of the form
$e^{i\gamma\eta_x\eta_y/l^2}$. Thus, the ground state at any $\alpha$
can be written for any ${\alpha^{CF}}$, and the ground state energy is
not useful in selecting the optimal ${\alpha^{CF}}$.

To find the optimal CF anisotropy ${\alpha^{CF}}$ we turn to the
equal-time single-particle correlator $G(x,y)$ which is an implicit
function of $\Delta_{mn}$.  It is easily checked that $G(x,y)$ is
(nearly) independent (for large but finite $n_{max}$) of
${\alpha^{CF}}$ in the same way that the $E_{gs}$ per particle is (see appendix).

Given a $\alpha$, we will define the optimal value of
${\alpha^{CF}_0}(\alpha)$ by demanding that
$G(\frac{x}{{{\alpha^{CF}_0}}},y{{{\alpha^{CF}_0}}})$ be as close to
isotropic as possible. Fig. \ref{correlator37} shows the unscaled
correlators in the two directions $G(r,0)$ and $G(0,r)$ versus $r$, as
well as the scaled versions with the optimal scaling
($\alpha^{CF}_0=1.46$) for $\nu=\frac{3}{7}$ in the LLL for the
Coulomb interaction. As can be seen, one cannot make the correlators
in the $x,\ y$ directions truly equal by a simple rescaling. The best
that can be done is to arrange for the maxima and minima to be at the
same location.

\begin{figure}
\includegraphics[scale=0.35]{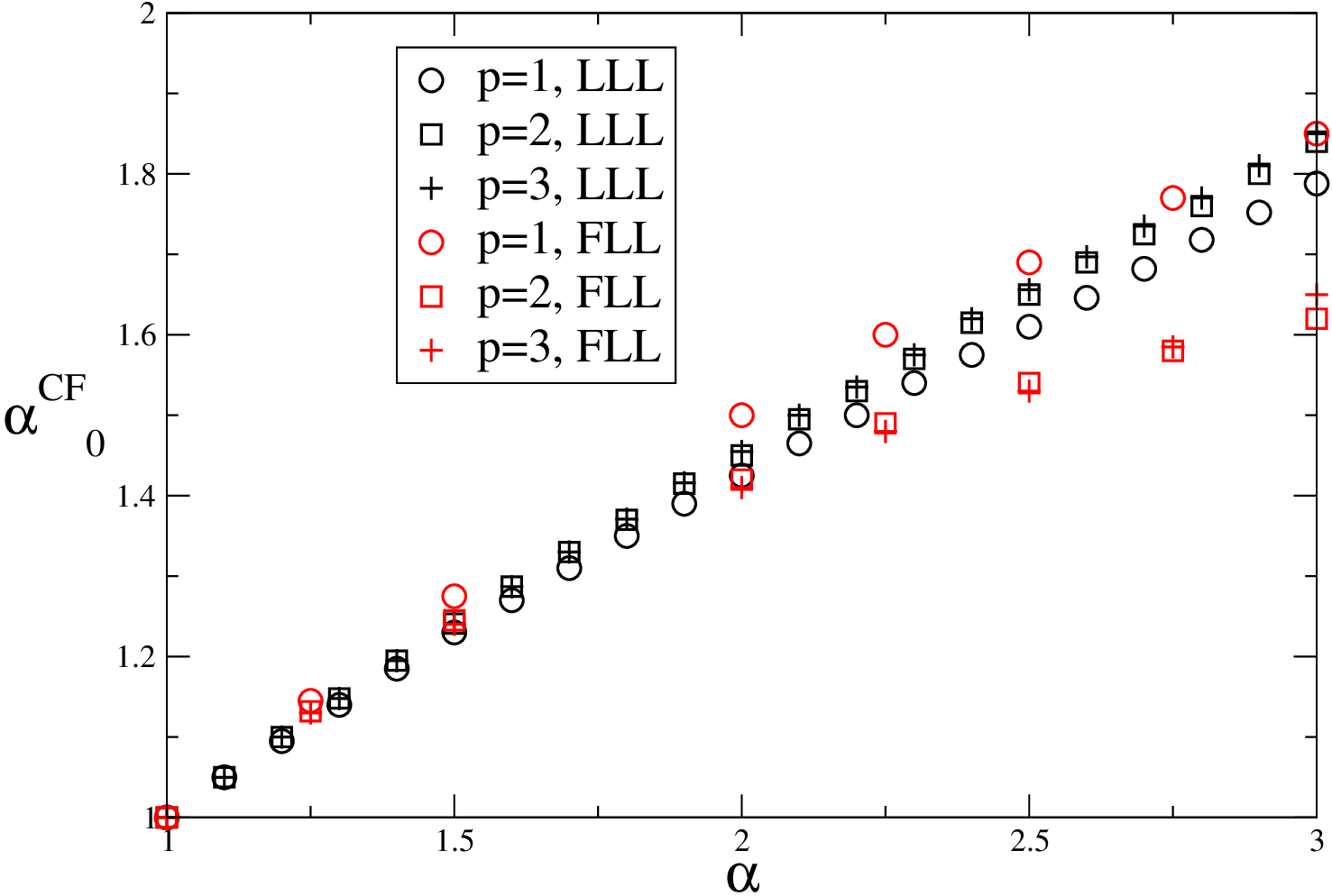}
\caption{The optimal value of ${\alpha^{CF}}$ plotted as a function of $\alpha$ for 
the first few principal fractions in the LLL and the FLL. The size of
the symbols roughly captures the uncertainty of estimating $\alpha^{CF}_0$
by trying to collapse $G(r/\alpha^{CF}_0,0)$ and $G(0,r\alpha^{CF}_0)$. }
\label{zeta0}
\end{figure}

In this way we determine $\alpha^{CF}_0(\alpha)$, which depends on the
interaction (Coulomb), the LL into which we have projected the
electron density, and the principal fraction in question (represented
by the integer $p$, where $\nu=\frac{p}{2p+1}$). As $p$ becomes larger
$\alpha^{CF}_0$ saturates to a value we assume is the relevant one for
$\nu=\half$.  Fig. \ref{zeta0} shows the dependence for the LLL and the first
Landau level (FLL).
For $\nu=\third$ in the LLL, our result is close to that of Yang {\it
et al}\cite{variational-Haldane} who estimate $\alpha^{CF}_0$ by
finding the largest overlap of the exact ground state with an
anisotropic Laughlin wave function\cite{Haldane-geo}. Their definition
of the anisotropy is the square of ours. Looking Figure 3 of Yang {\it
et al}\cite{variational-Haldane} and translating to our anisotropies,
we see that at $\alpha=\sqrt{2}$ they get
$\alpha^{CF}_0=\sqrt{1.37}=1.17$, while our result is
$\alpha^{CF}_0=1.185$.  

Another criterion for the optimal $\alpha^{CF}$ might be to determine
when the state is closest to the correponding isotropic Jain
state\cite{jain}. One can compute the total occupation of the lowest
$p$ CFLLs as a function of $\alpha^{CF}$ and find $\alpha^{CF}_0$ by
maximizing it. Reassuringly, this criterion gives the same
$\alpha^{CF}_0$ as does the collapse of $G(x
\alpha^{CF}_0,0)$ and $G(0,y\alpha^{CF}_0)$ (see Appendix). 

Finally, our results for the exactly solvable gaussian interaction in
the LLL $v(q)=e^{-(\lambda ql)^2/2}$ agree perfectly (see Appendix) with
the expression of Kun Yang\cite{kunyang}
$\alpha^{CF}_0=\big(\frac{\alpha^2+\lambda^2}{\alpha^{-2}+\lambda^2}\big)^{1/4}$
for all fractions. These checks show that CFHF captures the essential
physics of the anisotropic FQHE.

It is interesting to see that in the LLL, $\third$ has a higher value
of $\alpha^{CF}_0$ for a given $\alpha$ than other principal fractions, while
the reverse is true in the FLL. For a given $\alpha$, as $p$ increases
$\alpha^{CF}_0$ saturates rapidly. Experimental measurements of the
CF anisotropy have been carried out at $\nu=\half$ by Kamburov {\it et
al }\cite{Kamburov1}. Noting that their definition of the anisotropy
is the square of ours, they find at $\alpha=\sqrt{3}$ a CF-anisotropy of
$\alpha^{CF}_0=\sqrt{1.3}\simeq1.15$\cite{Kamburov1}.

\begin{figure}
\includegraphics[scale=0.35]{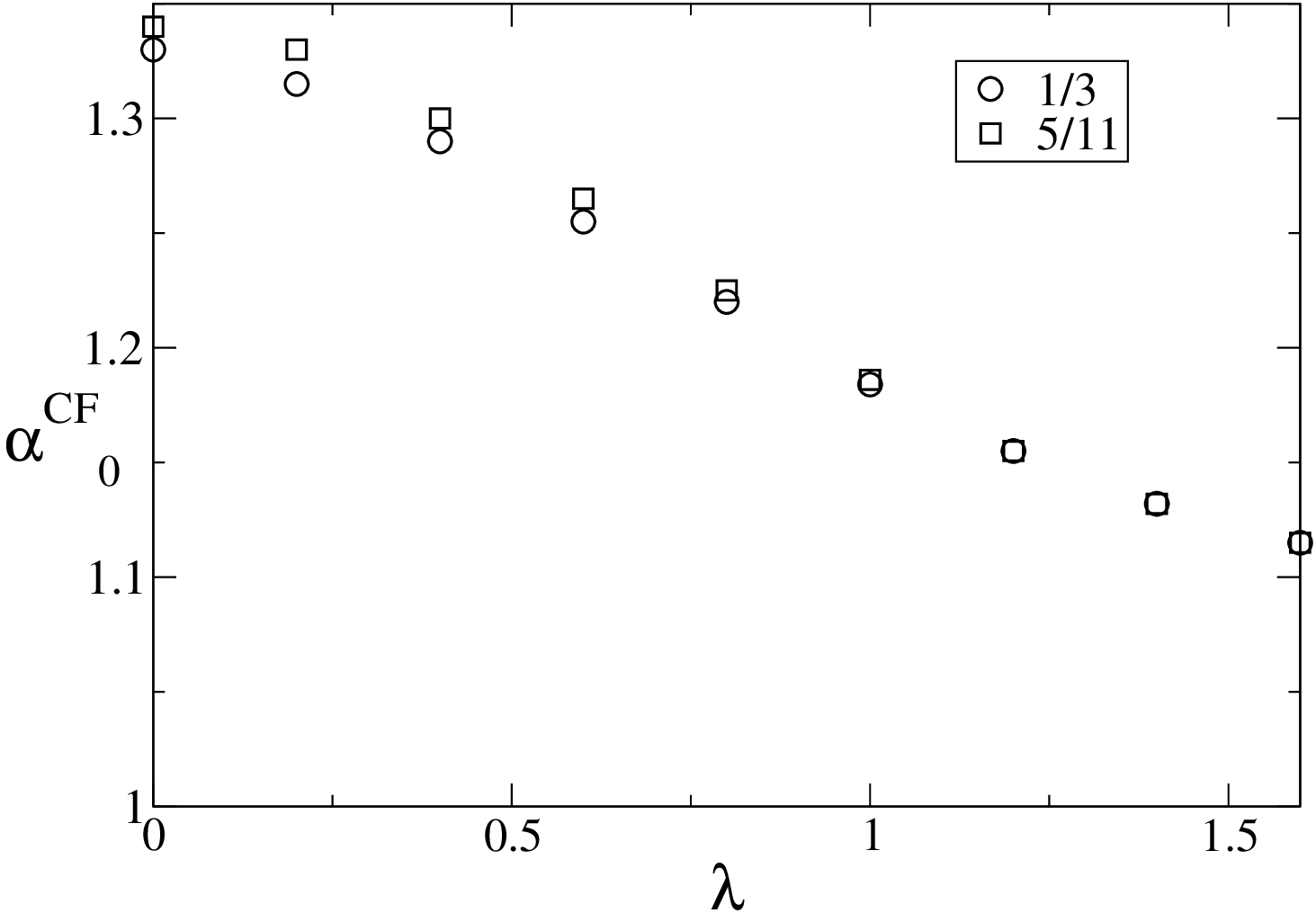}
\caption{The variation of $\alpha^{CF}_0$ with the parameter $\lambda$ modelling sample 
thickness ina Coulomb interaction for $\third$ and $\frac{5}{11}$. The
latter fraction is a proxy for the $\nu=\half$ state. }
\label{thickness}
\end{figure}

For the Coulomb interaction, for $\alpha=\sqrt{2}$ and large
$p=3,4,5$, we find a $\alpha^{CF}_0$ of about $1.34$, which is high
compared to the experiment. To understand the discrepancy, we turn to
a cartoon model of the finite thickness of the 2DEG, and its effects
on $\alpha^{CF}_0$. We will assume that the Coulomb potential has been
modified to

\beq
v_{\lambda}(q)=\frac{2\pi e^2}{\varepsilon q}e^{-(\lambda q l)^2/2}
\eeq
where $\lambda$ is a dimensionless parameter representing the
thickness of the 2DEG. A reasonable value of $\lambda$ for an
experimental sample would be of order 1. Fig. \ref{thickness} shows
the variation of $\alpha^{CF}_0$ with $\lambda$ for $\alpha=\sqrt{3}$.

As expected, $\alpha^{CF}_0$ decreases with thickness, and reasonable
values of $\lambda\simeq1$ can give rise to the experimentally
measured CF anisotropy\cite{Kamburov1}.

In summary, we have generalized the Hamiltonian approach for the FQH
regime derived by R. Shankar and the present
author\cite{prl-us,rmp-us} to the case of systems with electronic
anisotropy. This approach applies to any Landau level, arbitrary
interactions, and any principal fraction. While the exact spectrum of
the Hamiltonian in the enlarged Hilbert space is identical to that of
the electronic Hamiltonian, the usefulness of the approach lies in
approximations such as Composite Fermion Hartree Fock. We find that
the equal time single-CF correlator provides a good way to estimate
the CF anisotropy $\alpha^{CF}_0$ for any given $\alpha$. Our results
are in good to excellent agreement with previous
ones\cite{variational-Haldane,kunyang}. For principal fractions
labelled by $p$, where $\nu=\frac{p}{2p+1}$, $\alpha^{CF}_0$ increases
with $p$ in the LLL, but decreases with $p$ in the first Landau
level. In both cases $\alpha^{CF}_0$ saturates rapidly with increasing
$p$.

We have also investigated the effect of finite sample thickness by
using a model potential. We find that $\alpha^{CF}_0$ decreases with
sample thickness, and reasonable values of the thickness parameter
give CF anisotropies in agreement with experiment\cite{Kamburov1}.

There are many directions in which this approach can be extended. The
conserving approximation\cite{resp-fns-us} can be used to compute
magnetoexciton dispersions and response functions. This will help us
identify potential instabilities of anisotropic FQH states. Recall
that we restricted our variational search to translation invariant
states. It is possible that stripe states formed of CFs (natural in
the presence of anisotropy and CFLL mixing) are lower in energy than
translationally invariant ones. Such states, when suitably dressed by
quantum/thermal fluctuations and disorder may help us understand the
peculiar behavior of the longitudinal resistivities at
$\nu=\frac{7}{3}$\cite{xia-73}, which is currently not understood
(see, however, \cite{mulligan}).

A second direction in which this approach could be useful is in
analyzing anisotropic strongly correlated states in topological
(Chern) bands\cite{chern} and two-dimensional time-reversal invariant
topological insulators (2DTIs)\cite{2dti}. There has been much recent
excitement with the numerical discovery of FQH states in such
systems\cite{2dti-numerics}. For topological bands and 2DTIs with
conserved $S_z$, R. Shankar and the author have shown\cite{TI-us} that
the interacting Hamiltonian can once again be analyzed in CF
language. The author intends to investigate these and other issues in
the near future.

The author is grateful to the Aspen Center for Physics (NSF 1066293)
for its hospitality while this work was conceived. He would also like
to thank J. K. Jain, R. Shankar, and H. A. Fertig for illuminating
conversations, and is grateful for partial support from NSF-PHY
0970069 (GM), and the US-Israel Binational Science Foundation-2012120.

\section{Appendix A}
\begin{figure}
\vspace{0.25in}
\includegraphics[scale=0.3]{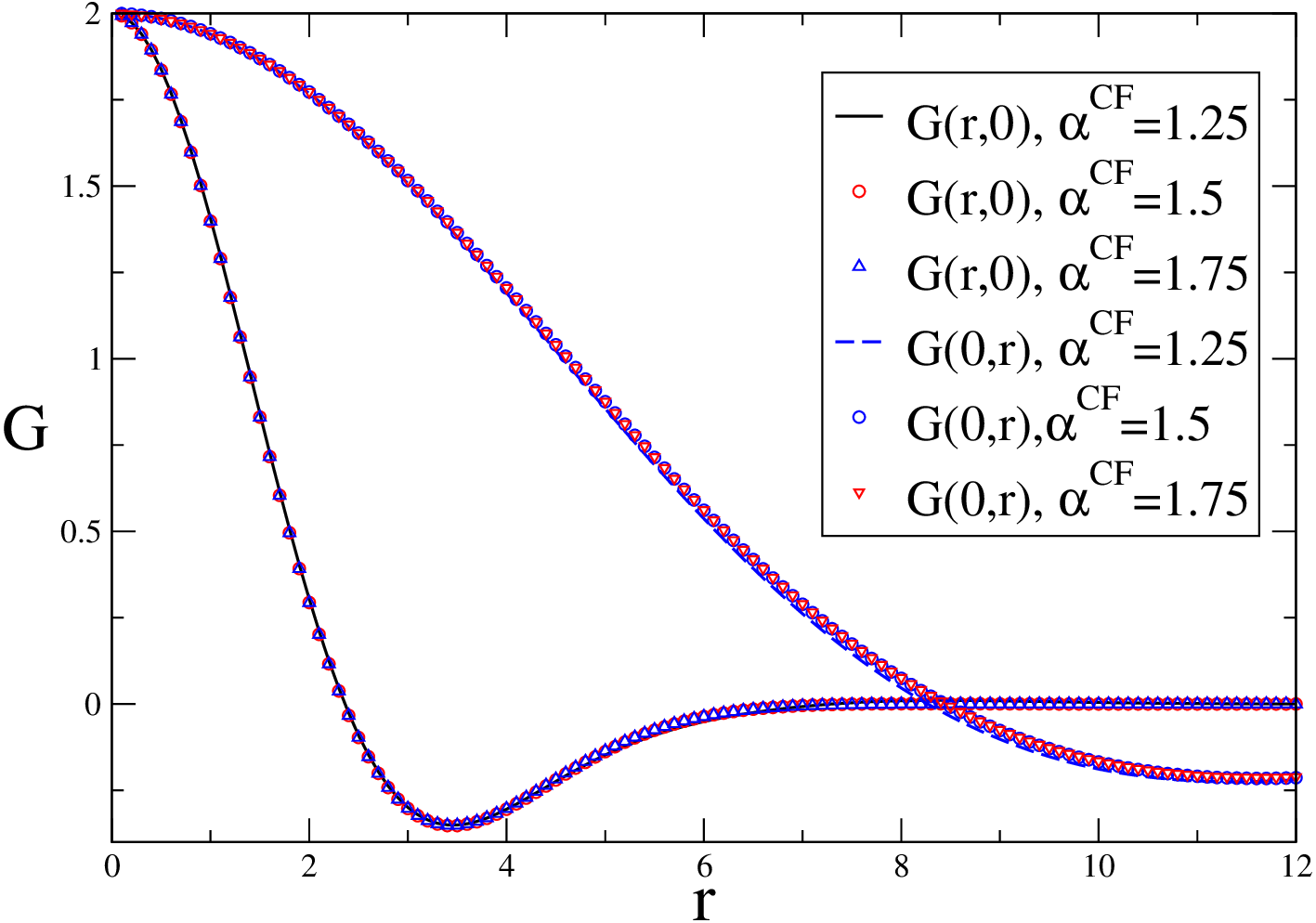}
\caption{The unscaled correlators $G(r,0)$ and $G(0,r)$ for $\alpha=3$
  and $\nu=\frac{2}{5}$ for different values of $\alpha^{CF}$. The
  interaction is Coulomb. }
\label{alpha-independence}
\end{figure}

In order to use the equal-time correlator to extract the effective CF
anisotropy $\alpha^{CF}_0$, we need to be sure that the correlator is
independent of the nominal value of $\alpha^{CF}$. Fig. \ref{alpha-independence} shows the
$\alpha^{CF}$-independence of the unscaled correlators for
$\nu=\frac{2}{5}$. The electronic anisotropy is $\alpha=3$, the
nominal values of $\alpha^{CF}_0$ run from 1.25 to 1.75, and 9 CFLLs
are being kept. This holds true of arbitrary fractions and any
$\alpha$.

\begin{figure}
\vspace{0.5in}
\includegraphics[scale=0.3]{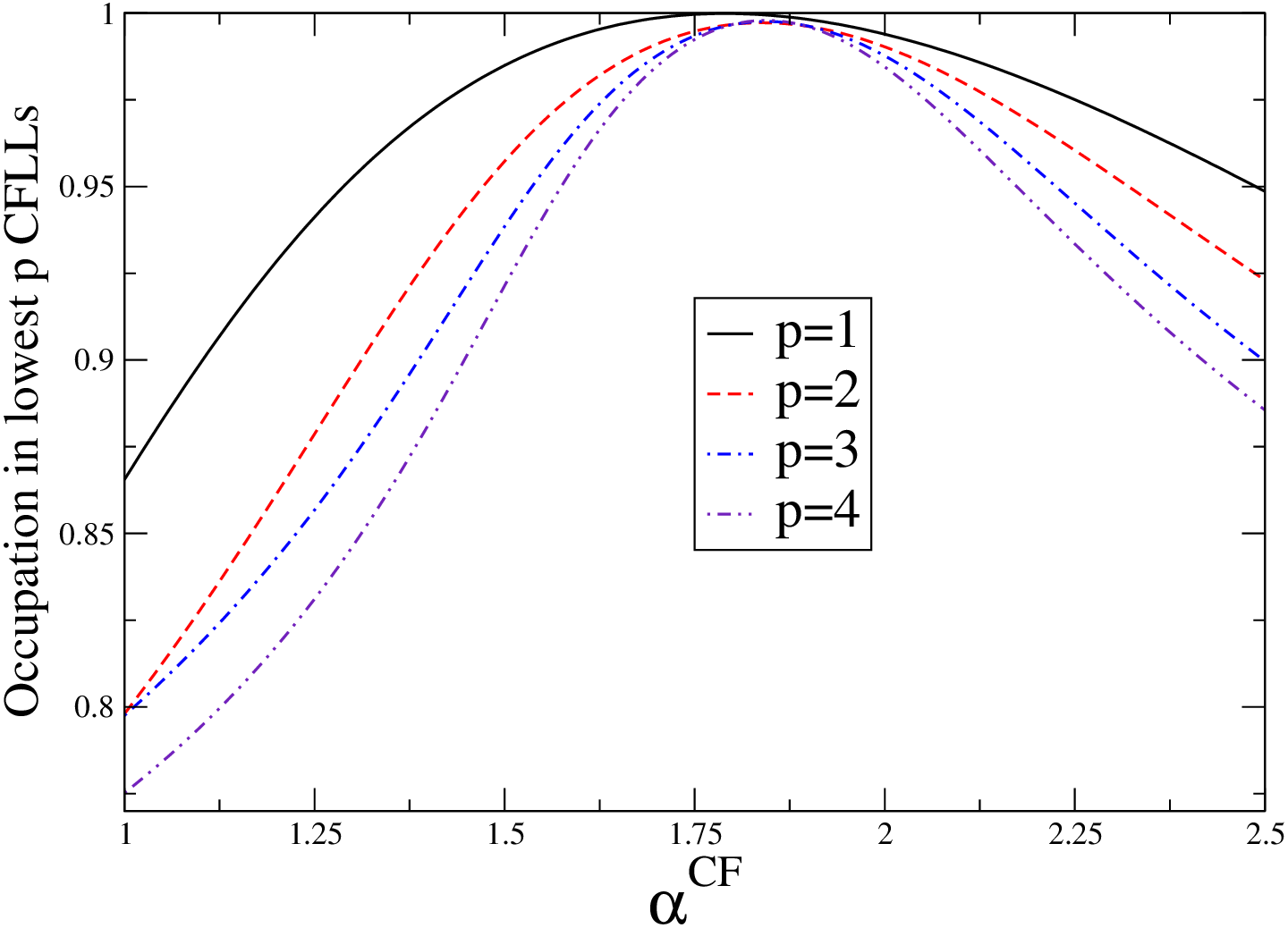}
\caption{The normalized occupation of the lowest $p$ CFLLs versus $\alpha^{CF}$ for $\alpha=3$ in the LLL for the Coulomb interaction. }
\label{occLLL}
\end{figure}

\begin{figure}
\vspace{0.25in}
\includegraphics[scale=0.3]{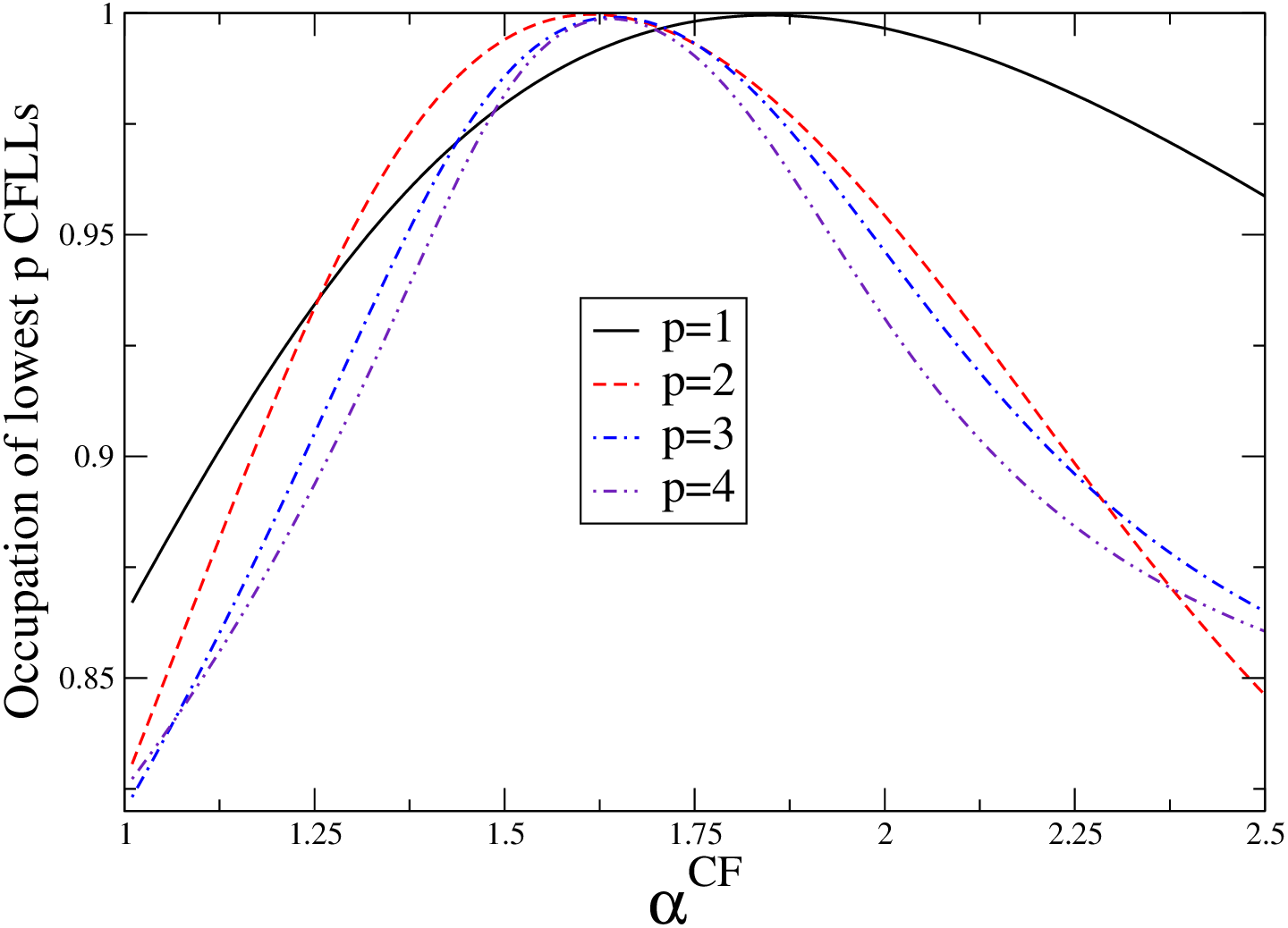}
\caption{The normalized occupation of the lowest $p$ CFLLs versus
  $\alpha^{CF}$ in the first Landau level for the Coulomb interaction.
}
\label{occFLL}
\end{figure}

A different way to extract the optimal CF anisotropy is to ask at what
$\alpha^{CF}$ the ground state is closest to the corresponding Jain
state of $p$ filled CFLLs\cite{jain}. Our state is in an enlarged Hilbert space,
so the best we can do is to compare the combined occupation of the
lowest $p$ CFLLs to $p$. Figs. \ref{occLLL} and \ref{occFLL} show the normalized occupation
$\frac{1}{p}\sum\limits_{0}^{p-1}\Delta_{nn}$ versus $\alpha^{CF}$ in
the LLL and FLL. 

Comparing to Fig. \ref{zeta0}, for each fraction, the value of
$\alpha^{CF}$ at which the normalized occupation is maximum is almost
identical to the $\alpha^{CF}_0$ obtained from the collapse of
$G(r/\alpha^{CF}_0,0)$ and $G(0,r\alpha^{CF}_0)$. This makes sense
because the Jain state with the lowest $p$ CFLLs filled is isotropic.

For the special case of a gaussian interaction ($v(q)=e^{-(\lambda
  ql)^2/2}$) in the LLL, $\alpha^{CF}_0$ can be exactly expressed in
terms of $\alpha$ and the length scale $\lambda$ of the
interaction as\cite{kunyang}
\beq
\alpha^{CF}_0=\big(\frac{\alpha^2+\lambda^2}{\alpha^{-2}+\lambda^2}\big)^{1/4}
\eeq

\begin{figure}
\vspace{0.25in}
\includegraphics[scale=0.3]{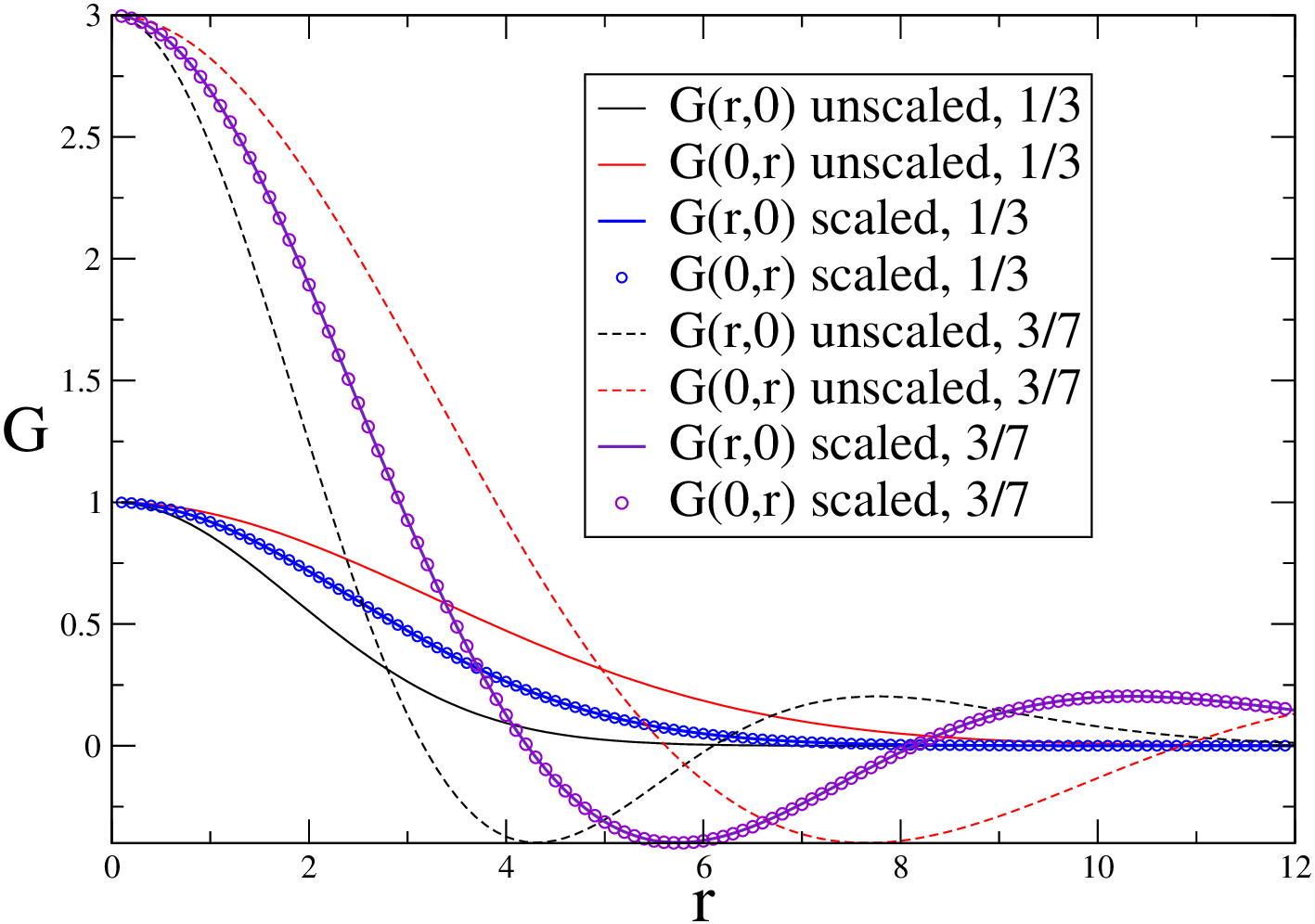}
\caption{The unscaled and scaled correlators for a pure gaussian
  interaction at $\alpha=3,\ \lambda=2$, for $\nu=\frac{1}{3}$ and $\frac{3}{7}$ in the LLL.}
\label{KY-corr}
\end{figure}

In Fig. \ref{KY-corr} we show the unscaled and scaled correlators in our
approach for $\frac{1}{3}$ and $\frac{3}{7}$, where the correlators have been scaled
with the exact value of $\alpha^{CF}_0$. The collapse of the correlator is nearly perfect. 

\end{document}